\documentclass{svproc}
\usepackage{url}
\usepackage{graphicx}
\usepackage{footmisc}
\usepackage [autostyle]{csquotes}
\usepackage{amsmath}

\begin{document}
\mainmatter              
\title{Dynamics of Wealth Inequality in Simple Artificial Societies}
\titlerunning{Wealth Inequality in Simple Societies}  
%
\author{John C. Stevenson}
\authorrunning{John C. Stevenson} 
%
%
\institute{Independent, Long Beach, NY, USA \\
\email{jcs@alumni.caltech.edu}}

\maketitle              

\begin{abstract}
	
A simple generative model of a foraging society generates significant wealth inequalities from identical agents on an equal opportunity landscape. These inequalities arise in both equilibrium and non-equilibrium regimes with some societies essentially never reaching equilibrium. Reproduction costs mitigate inequality beyond their affect on intrinsic growth rate. The highest levels of inequality are found during non-equilibrium regimes. Inequality in dynamic regimes is driven by factors different than those driving steady state inequality. Evolutionary pressures drive the intrinsic growth rate as high as possible, leading to a tragedy of the commons.

\end{abstract}
\section{Introduction}
%


Current studies on wealth inequality use many different approaches: stationary distributions \cite{benhabib}, geometric Brownian motion \cite{benisty}, models calibrated to actual economies \cite{berman}, minimal models of a system \cite{adam}, to cite just a few. Common to these modelling approaches are assumptions about the dynamics of the process, for example geometric Brownian motion (GBM) and random exchanges of assets \cite{bouchaud}. Since in GBM models the wealth distribution never reaches equilibrium and ends up concentrating all the wealth in a diminishing cohort \cite{adam}, mean reversion or wealth redistribution terms must be added \cite{berman}. Rather than assume processes and then fit to empirical data, this research builds a minimal generative model for a system that makes no assumptions about the underlying behaviors or processes other than the population-driven struggle for existence \cite{rough,gause}. The purpose is to investigate whether wealth inequality will emerge in societies with equal agents and equal opportunities, and if so, what drives these inequalities.

While this aggregation of simple individuals may barely qualify as a society or an economy, the agents interact with each other through the competition for resources (to survive) and space (to reproduce). From these interactions, complex behaviors of population-driven ecologies \cite{kot} and unequal wealth distributions emerge. These wealth distributions show inequalities that are dependent on the intrinsic growth rate of the population, the cost of reproduction, and whether the economy has reached equilibrium. Furthermore, by allowing natural selection to act at the individual level \cite{wilson}, the population responds with a classic tragedy of the commons. \cite{ostrom}


Epstein and Axtell's model \cite{eps:axl} is simplified by specifying identical agents on an equal opportunity (flat) landscape.\footnote{Epstein and Axtell \cite{eps:axl} [pgs. 32-37,122] detailed wealth inequalities and identified but did not pursue potential investigations. Others \cite{rahman,horres} considered more complex, evolving configurations without first addressing the underlying sensitivities and dynamics.} As in the study of bacteria \cite{gause}, the population trajectory begins with a single agent and, in these cases, is dependent on two growth characteristics, infertility and birth cost. Population trajectories from two constant parameter scenarios are used in the discussions of societies with growth in the stable regimes: one with \enquote{Low Fertility} and the other with significant \enquote{Birth Cost}. Two other trajectories with evolving parameter scenarios are discussed: one in a stable growth regime \enquote{Evolved Stable}, and one in a chaotic growth regime: \enquote{Evolved Chaotic}. Specific details for these four scenarios are provided in Table 1, and detailed descriptions of the parameters and processes of the simple artificial society model are given in Appendix A.

This simplified, population-driven model admits comparison with equation-based continuum modeling of single species populations, developed in the fields of mathematical biology and ecology \cite{kot,murray}. These comparisons validate the dynamics of the model, allow calculation of the intrinsic growth rate based on the family of Verhulst Processes \cite{verh,murray}, and define the various population level regimes of stable, oscillatory, and chaotic. Details of these comparisons and calculations are given in Appendix B. 

\begin{table}[h!]
	\begin{center}
	\begin{tabular}{|c|c|c|c|c|c|c|c|}
		\hline
		Scenario &  $f$ & $bc$ & $r$ & $K$ & $mA$ & $tW$ & $G$\\
		\hline
		Low Fertility & 85 & 0 & $0.018\pm0.4\%$ & $837\pm0.9\%$ & $473\pm9.7\%$ & $4767\pm4.7\%$ & $0.53\pm2.8\%$\\
		Birth Cost & 10 & 40 & $0.032\pm0.6\%$ & $824\pm0.4\%$ & $2264\pm3.2\%$& $16254\pm2.2\%$ & $0.33\pm2.9\%$\\
		Evolved Stable & 10 & 0 & $0.170\pm0.5\%$ & $859\pm3.9\%$ & $12.7\pm4.2\%$& $504\pm14\%$ & $0.688\pm3.9\%$\\
		Evolved Chaos & 1 & 0 & $1.542\pm14\%$ & $916\pm53\%$ & $1.41\pm19\%$& $480\pm79\%$ & $0.73\pm14\%$\\
		\hline
	\end{tabular}
\hfill \break
	\end{center}
	\caption{Artificial Society Scenarios}{Scenario simulation parameters infertility $f$, birth cost $bc$, the resultant intrinsic growth rate $r$, carry capacity $K$, mean age $mA$, total wealth $tW$, and Gini Coefficient $G$. Each scenario was run 100 times with different random seeds. Statistics were collected after the simulations had reached steady state (such as it is for the chaotic regime).}
\end{table}

\section{Wealth Inequality Dynamics}

Additional measures were used to develop distributions of the individuals' ages, surplus resources (wealth), and the deaths per cycle. These additional measurements enable a determination of steady state (equilibrium) and dynamic (non-equilibrium) conditions, allow detailed histories of individual and total wealth over time,\footnote{Comparisons of inequality distributions measured with a single ratio have both mathematical \cite{font,cowel} and practical difficulties \cite{weis,sitth}. Equally sized populations are fully sampled to address these difficulties. In addition, ratios only provide a relative measure of wealth so total wealth is included to address this issue.} and provide insight into both the relationship of mean age and mean death rate to the implied, intrinsic growth rate; and into the nature of the society's wealth.

\subsection{Dynamic relaxation times}
Figure 1(a) shows the elite (top 10\%) and overall mean age and Figure 1(b) shows the population level and Gini Coefficient; both figures are over time and are for the two constant parameter scenarios. While these populations reach and sustain their carry capacity levels quite quickly, achieving equilibrium, as measured by mean age and Gini Coefficient, takes much longer (for Low Fertility over 35,000 cycles, an impractical length of time). 

\begin{figure}
	\begin{center}
		\includegraphics[angle=-90,scale=0.65]{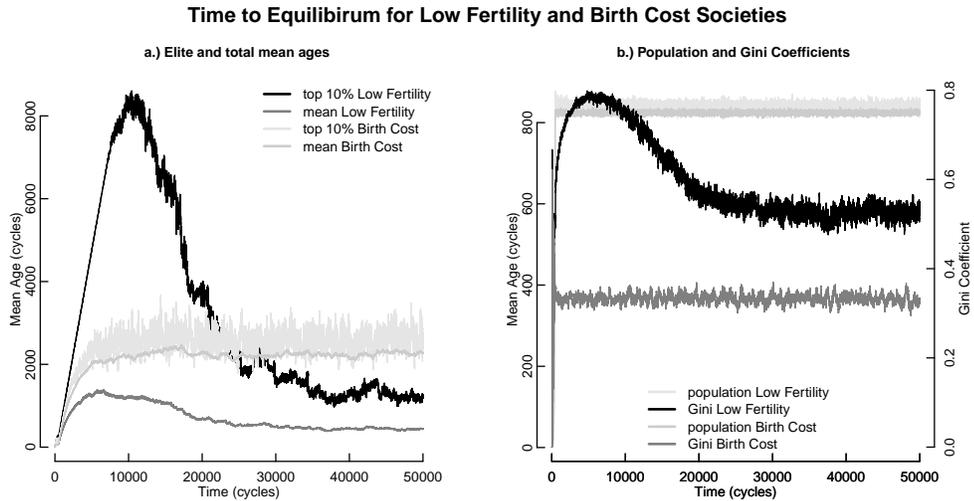}
	\end{center}
	\caption{Relaxation Times} {a). Elite (top 10\%) and overall mean ages and total wealth for the two constant parameter scenarios. b). Population level and Gini Coefficients for the same scenarios.}
\end{figure}

The initial agents (founders), born into a underpopulated and rich landscape, have a tremendous advantage in building up their personal wealth before the population reaches the actual carrying capacity. Once that carry capacity has been reached, equilibrium is not achieved until these founding agents have given up their surpluses and expired. Figure 2 shows the wealth histories for the early founders of these two societies. 

\begin{figure}
	\begin{center}
		\includegraphics[angle=-90,scale=0.65]{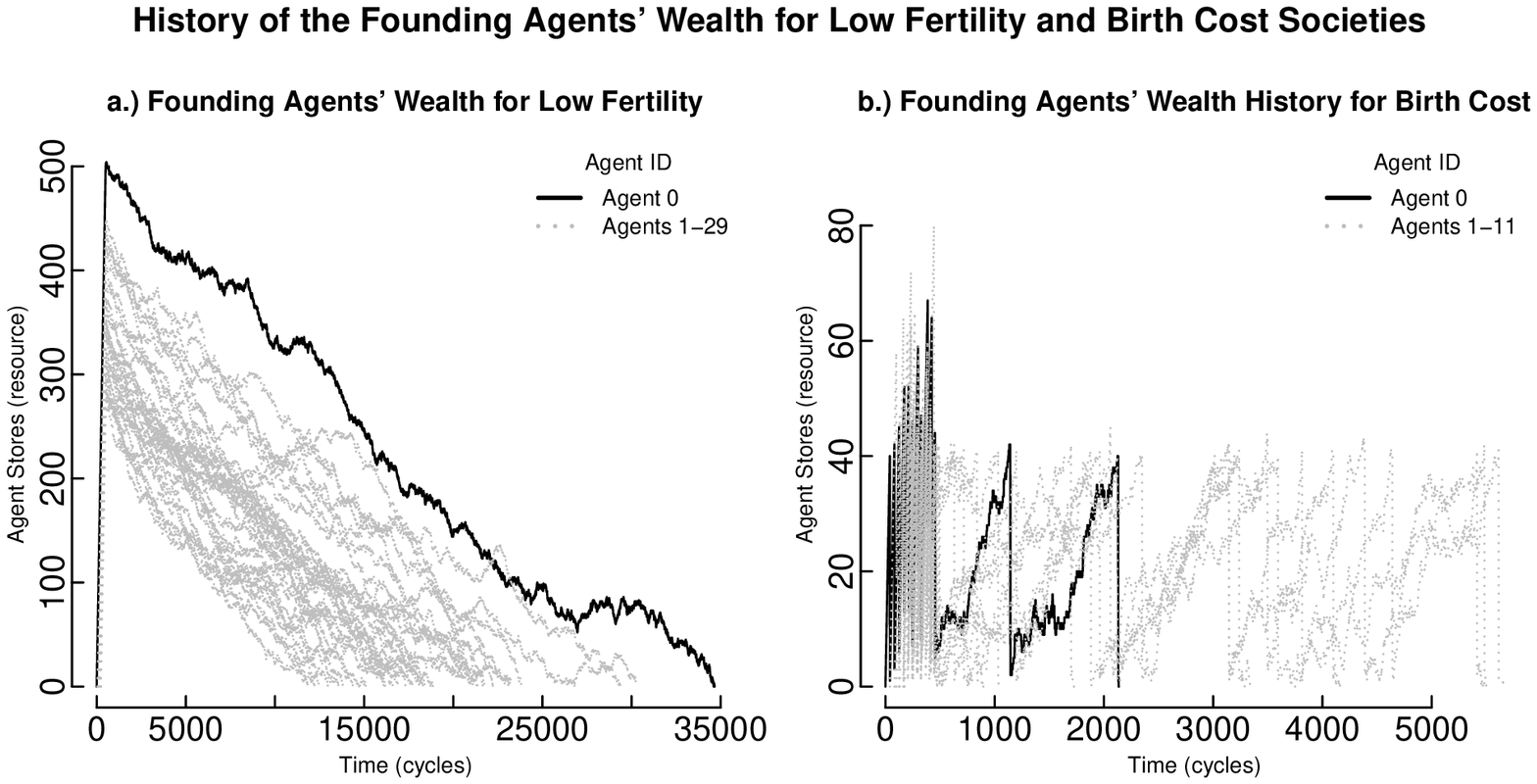}
	\end{center}
	\caption{Individual Wealth histories} {a.) Wealth history for the first thirty agents of the Low Fertility scenario.  b.)  Wealth history for the first twelve agents of the Birth Cost scenario.}
\end{figure}

The relationships of inequality and mean age to the growth parameters are substantially different between a society at steady state versus one still in transition.

\subsection{Equilibirum Inequality and Sensitivities}
While Low Fertility highlights the lack of equilibrium, Birth Cost shows that even societies that do attain equilibrium in a reasonable time have significantly unequal wealth distributions. Figure 3(a) shows, at equilibrium, the relationship of total wealth to mean age (a proxy for intrinsic growth rate). Surprisingly even with  increasing birth costs, which are sunk costs, the total wealth compares favorably. Figure 3(b) shows decreasing inequality (Gini Coefficient) with increasing mean age. Two very different populations emerge with similar mean ages but significantly different inequality measures. (This effect can also be seen in Table 1.) Also, as mean age decreases to and below 10, the effects of chaotic population trajectories appear.

\begin{figure}
	\begin{center}
	\includegraphics[angle=-90,scale=0.65]{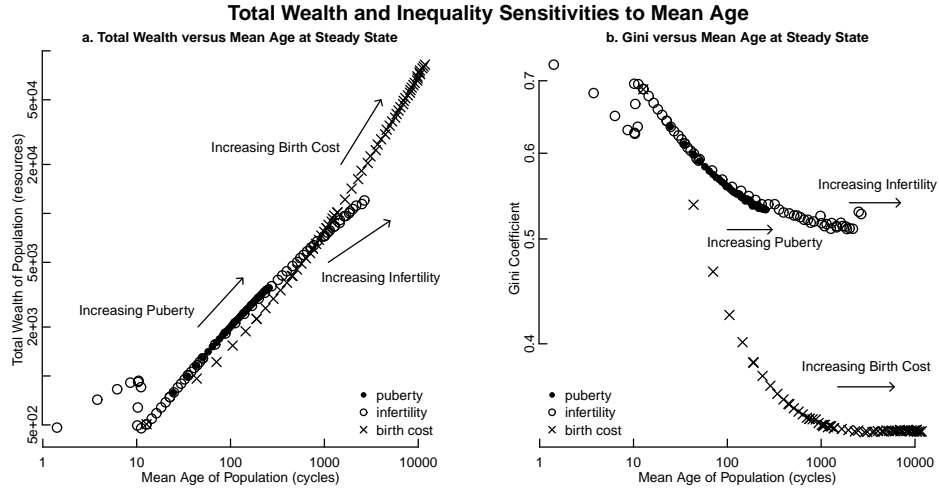}
	\end{center}
	\caption{Total Wealth and Inequality at Steady State} {a) The relationship of population's total wealth to mean age. b.) The relationship of the Gini Coefficient to mean age. Note the strikingly different inequality for the same mean ages.}
\end{figure}

\subsection{Natural Selection of Growth Parameters}
By applying natural selection pressures to the infertility and birth cost parameters, the artificial society transitions from a complex system to a complex adaptive system \cite{wilson}. Since the selection is occurring on the individual level, it's classified as a within-group selection. Figure 4 shows that from the initial uniform distribution of infertility and birth cost parameters across the broad ranges allowed, evolution quickly and forcefully selects for the highest possible intrinsic growth rate. This selection results in the lowest mean ages, the lowest total wealth and the highest wealth inequality of all the scenarios at steady state. Even with the range of these parameters restricted to the regimes of stable population levels, the evolution of the agents' reproductive parameters firmly demonstrate a tragedy of the commons so often associated with with-in group selection. Table 1 provides the total wealth and wealth inequality measures for all the scenarios for comparison, which highlights the carnage of the within-group evolution. The generation of endogenous between-group selection, from which cooperation can emerge and the tragedy of the commons avoided, is the next challenge for these simple artificial societies. \cite{wilson,ostrom,pepper,tverskoi}.

\begin{figure}
	\begin{center}
		\includegraphics[angle=-90,scale=0.65]{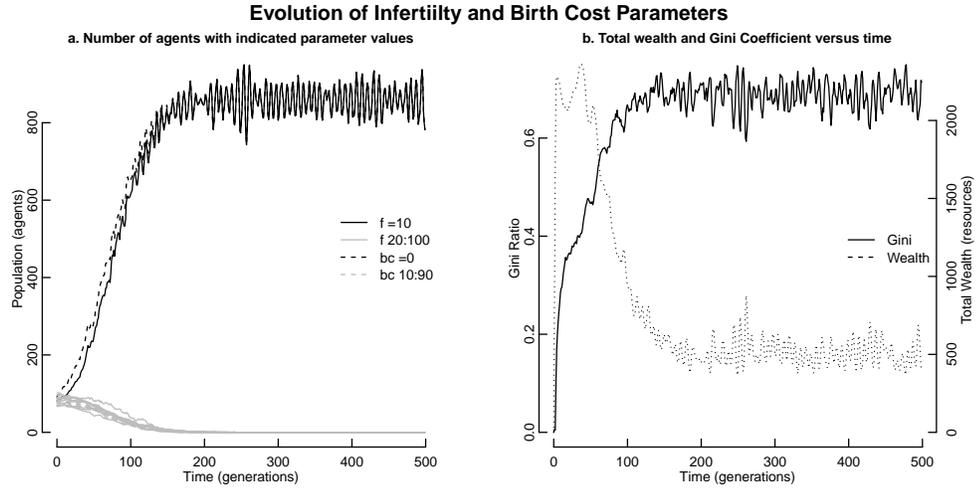}
	\end{center}
	\caption{Natural Selection of Reproductive Parameters} {a) The selection of infertility and birth cost parameters from a uniform initial distribution limited to the stable population level regimes. b.) The trajectories of the mean ages and total wealth of the society undergoing this natural selection.}
\end{figure}

\section{Conclusions}

Simple societies with equal opportunity environments and equally capable individuals generate complex wealth distributions whose inequalities are dependent on the intrinsic growth rate of the population, the cost of reproduction, and whether the society has reached equilibrium. Some societies never achieve equilibrium in a reasonable time. Determination of the relaxation times of wealth distributions is of interest in modern and complexity economics \cite{rosser,wilsonKirman}.

The drivers of these inequalities are shown to be much different in the steady state phase than during non-equilibrium transitions. The degree of inequality was shown to be lower and the total wealth higher with slower intrinsic growth rates (lower death rates) at steady state. The inclusion of birth costs made additional contributions to reduced inequality beyond its effects of reducing intrinsic growth while actually increasing the total wealth of the population even though these resources were consumed. It is clear that under most configurations, significant inequalities persist even though agents' capabilities and resource opportunities are equal.

After a population has achieved a true steady state, increasing birth cost and infertility decrease the implied, intrinsic growth rate. As the intrinsic growth rate decreases, mean ages and total wealth increase while mean deaths per cycle and inequality decrease. These relationships suggest that the larger inequalities at steady state are characteristic of short-lived populations driven by high death rates. Increasing birth costs reduce inequality at a much greater rate than increasing infertility. And it is perhaps counter-intuitive that these increasing birth costs do not reduce the total wealth of the population though they represent a sunk cost of wealth. Surprisingly, the Birth Cost simulation with the most equal society also has the largest total wealth. The implications for policy would be significant if these birth cost effects are found in natural societies.

The highest levels of inequalities are found in non-equilibrium periods during and after the initial growth phase, when a small number of agents (founders) reproduce slowly into a rich, under-populated landscape. These founders store such significant resources before the population reaches its carry capacity that even after the carry capacity has been reached, the residual surplus resources decay quite slowly, preserving high inequality and preventing equilibrium for time periods orders of magnitude greater than the scale of the initial growth phase. Though these results are for a minimal model of a system, the parallels to the fortunes built at the beginnings of technological and social revolutions are inescapable.

For these simple forging societies, natural selection drives their reproductive parameters to those values which maximize the individuals' reproductive rates (intrinsic growth rates). While this selection pressure at the individual level successfully maximizes the society's intrinsic growth rate, this selection had a devastating effect on the society's total wealth and stability while generating the highest wealth inequalities and lowest mean ages seen in any of these scenarios at steady state. Truly a tragedy of the commons and one not unfamiliar to human societies. 

It is surprising and encouraging to see that even from such simple artificial societies with equal agents given equal opportunities, wealth inequalities emerge in dynamic (founders effects), equilibrium (including the out-sized impact of birth costs), and evolving (tragedy of the commons) conditions.  

%
%

\appendix{\textbf{Appendix A - Computational Model and Process}}

\section{Model Parameters}
Table 2 provides the definition of the agents' and landscape's parameters for this simple model. Vision and movement are along rows and columns only. The two dimensional landscape wraps around the edges (often likened to a torus). Agents are selected for action in random order each cycle. The selected agent moves to the closest visible cell with the most resources with ties resolved randomly. After movement, the agent harvests and consumes (metabolizes) the required resources. At this point, if the agent's resources are depleted, the agent is removed from the landscape. Otherwise an agent of sufficient age (puberty) then considers reproduction, requiring sufficient resources (birth cost), a lucky roll of the fertility die (infertility), and an empty von Neumann neighboring cell. (The von Nuemann neighborhood consists of only the four neighboring spaces one step away by row or column.) If a birth occurs in a configuration with zero puberty, the newborn is added to the list of agents to be processed in this current cycle. Otherwise (puberty $>0$), the newborn is placed in the empty cell and remains inert until the next action cycle. With this approach for the action cycle, no endowments are required either for new births or for the agent(s) at start-up. Once all the agents have cycled through, the landscape replenishes at the growth rate and the cycle ends.

\begin{table}[h!]
	\begin{center}
	\begin{tabular}{|c|c|c|c|c|}
		\hline
		Agent Characteristic & Notation & Value & Units & Purpose \\
		\hline
		vision & $v$ &  6 &  cells & vision of resources on landscape \\
		movement & -- &  6 &  cells per cycle &  movement about landscape \\
		metabolism & $m$ & 3 & resources per cycle &  consumption of resources \\
		birth cost & $bc$ & 0,40 & resources &  sunk cost for reproduction \\
		infertility & $f$ & 10,85 & 1/probability & likelihood of birth \\
		puberty & $p$ & 1 &  cycles &  age to start reproduction\\
		surplus & $S$ & 0+ & resources &  storage of resources across cycles \\
		\hline
	\end{tabular}
	\bigbreak
	\begin{tabular}{|c|c|c|c|}
		\hline
		Landscape Characteristic & Notation & Value & Units\\
		\hline
		rows & -- & 50 & cells \\
		columns & -- & 50 & cells\\
		max capacity &$R$ & 4 & resources per cell\\
		growth & $g$ & 1 & resources per cycle per cell \\
		initial & $R_{0}$ & 4 & resources, all cells\\
		\hline
	\end{tabular}
	\caption{Agent and Landscape Parameters of the Simple Economic Model}
	\label{Table 1:}
	\end{center}
\end{table}

One metabolism rate ($m$), uniform across a given population, is modelled and takes the value that consumes, per cycle, 25\% less than the maximum capacity per cell. This value allows the agent to gather more resources than those required by its per cycle metabolism and is called a surplus society in contrast to a subsistence society where the maximum resources in a cell are equal to the metabolism of the agent. And, again for simplicity, the vision and movement characteristics are set to equal values of distance.
\section{Conservation of Resources}
The calculation of conservation of energy (resources) confirms the validity of the simulation, provides a precise description of the computational process, and provides independent measurements of the internal resource flows. Figure 5 details the control volumes used for this analysis of the conservation of resources. The source is growth of resources in the landscape and the sinks are agent metabolism, death, and birth costs (if any). 

The landscape resource conservation equation can be written as:
\begin{equation}
	\Delta E_{L}(t)= \sum_{c=1}^{N_{c}}g_{c}(t-1)-F(t)
\end{equation}
where $\Delta E_{L}$ is the change in total resources of the landscape from the end of the previous cycle $t-1$ to the current cycle $t$, $F(t)$ are the resources foraged by the agents, $N_{c}$ is the number of cells in the landscape, $g_{c}(t-1)$ is the resource added to cell $c$ at the end of the previous cycle. $g_{c}(t)$ is given as:
\begin{equation}
	g_{c}(t) =  
	\begin{cases}
		g & r[c,t]+g\le R\\
		R-r[c,t] & r[c,t]+g>R\\
	\end{cases}
\end{equation}	 
where $R$ is the maximum resources in a landscape's cell, and $g$ is the growth rate of resources in landscape's cell. The resources foraged $F(t)$ from the landscape by the agents is defined as:
\begin{equation}
	F(t) = \sum_{a=1}^{A(t-1)}r[c(a),t]+\delta_{p_{0}}\sum_{a=1}^{B(t)}r[c(a),t]
\end{equation}
where $a$ is the agent index, $A(t-1)$ is the list of agents alive at the end of the previous cycle, $B(t)$ is the list of new agents generated in this cycle, $\delta_{p_{0}}$ is one if the puberty parameter equals 0 and 0 otherwise, $c(a)$ is the cell location of an agent indexed as $a$, and $r[c(a),t]$ are the resources in the cell occupied by $a$ that are foraged by $a$ at the current time.

The conservation equation for the change in resources of the population $\Delta E_{P}(t)$ from the previous to the current cycle can now be written as:
\begin{equation}
	\Delta E_{P}(t) = F(t)-\sum_{a=1}^{A(t)}m -\sum_{a=1}^{D(t)}(S_{a}(t)+m)-\sum_{a=1}^{B(t)}[bc -\delta_{p_{0}}m]
\end{equation}
where $A(t)$ is the list of agents alive, $m$ is the (constant) metabolism, $bc$ is the (constant) birth cost, $D(t)$ is the list of agents that died in this cycle and $S_{a}(t)$ is the surplus resources of those agents $a$ on list $D$ which have died ($S_{a}(t)<0$) in this cycle so that $S_{a}(t)+m$ are the (positive) resources lost upon its death.
\\

\appendix{\textbf{Appendix B - Single Species Models from Mathematical Biology}}
\\

A continuous homogeneous model of a single species population $N(t)$ was proposed by Verhulst in 1838 \cite{verh} :

\begin{equation}
	\frac{dN(t)}{dt}=rN(1-\frac{N}{K})
\end{equation}
where $K$ is the steady state carry capacity, $t$ is time, and $r$ is the intrinsic rate of growth. This model represents self-limiting, logistic growth of a population and is used to estimate the intrinsic growth rates $r$ of the population trajectories generated by this simple model.

A discrete form of the Verhulst process incorporating an explicit time delay $\tau$ in the self-limiting term was proposed by Hutchinson \cite{hutch} to account for delays seen in animal populations. The resulting discrete-delayed logistic equation, often referred to as the Hutchinson-Wright equation \cite{kot}, is then

\begin{equation}
	N(t+1)=[1+r-\frac{N(t-\tau)}{K}]N(t)
\end{equation}
This model's intrinsic growth rate with $\tau = 3$ captures the steady state, oscillating, and chaotic populations trajectories seen in the simple model with similar intrinsic growth rates. This delay term represents the number of generations a landscape cell needs to restore its resources to the metabolic requirement of the agents. Figure 6(a) provides sample population trajectories for the simple model with implied intrinsic growth rates and Figure 6(b) displays discrete Verhulst process trajectories at critical intrinsic growth rates. These figures demonstrates the relationship of the various growth regimes to their respective intrinsic growth rates.

\begin{figure}
	\begin{center}
		\includegraphics[angle=-90,scale=0.65]{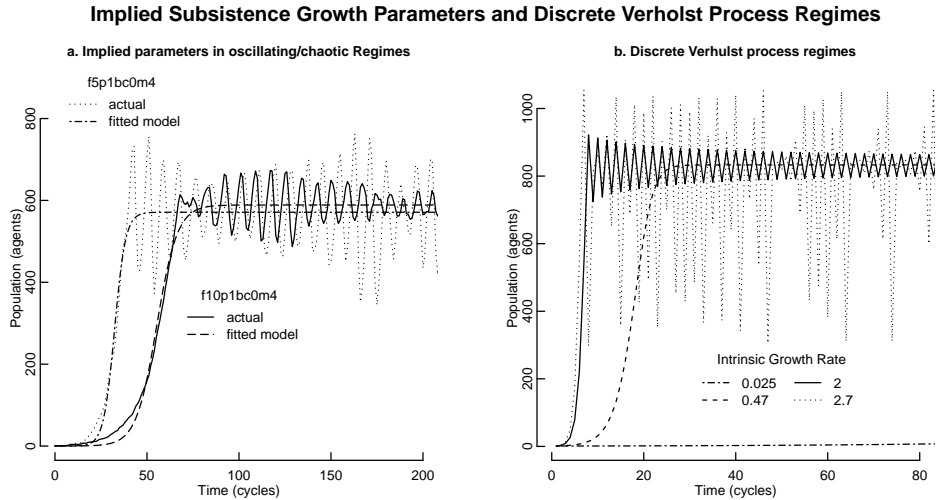}
	\end{center}
	\caption{a) Implied growth parameters for representative oscillating/chaotic, simple subsistence societies. f5p1bc0m4 is a society with $f$ of 5, puberty of 1, $bc$ of 0, and $m$ of 4. f10p1bc0m4 has a $f$ of 10 and the remaining parameters the same as before. b) Representative critical regimes of the discrete Verhulst process for simple surplus societies.} 
\end{figure}

\end{document}